\documentclass[aps,prl,twocolumn,superscriptaddress,amsmath,amssymb]{revtex4-2}

\usepackage{graphicx}
\usepackage{dcolumn}
\usepackage{bm}
\usepackage[T1]{fontenc}
\usepackage[english]{babel}
\usepackage{xcolor}
\usepackage{amsmath}

\begin{document}

\title{DNA handles bias force-dependent looping times}

\author{Wout Laeremans}
\affiliation{Soft Matter and Biological Physics, Department of Applied Physics
and Science Education, and Institute for Complex Molecular Systems,
Eindhoven University of Technology, P.O.\ Box 513, 5600\,MB Eindhoven,
Netherlands}

\author{Jef Hooyberghs}
\affiliation{UHasselt, Faculty of Sciences, Data Science Institute,
Theory Lab, Agoralaan, 3590 Diepenbeek, Belgium}

\author{Wouter G. Ellenbroek}
\email{w.g.ellenbroek@tue.nl}
\affiliation{Soft Matter and Biological Physics, Department of Applied Physics
and Science Education, and Institute for Complex Molecular Systems,
Eindhoven University of Technology, P.O.\ Box 513, 5600\,MB Eindhoven,
Netherlands}

\date{\today}

\begin{abstract}
DNA loop formation is central to gene regulation, and its kinetics depend sensitively on mechanical tension. In both single-molecule experiments and biological settings, this tension is typically transmitted through flanking DNA segments rather than acting directly at the looping sites. Using molecular dynamics simulations of the wormlike chain model, we show that such flanking segments substantially enhance the force dependence of looping, with the effect saturating once their length reaches approximately one DNA persistence length and disappearing when the connection is made flexible. We develop an analytic theory in which the handles extend the effective looping coordinate. This extension is set by two quantities: an equilibrium property of the construct that can be measured without force, and the force-dependent length over which a handle remembers the orientation of its binding site. The resulting theory quantitatively reproduces published optical-tweezer measurements, without adjustable parameters. Furthermore, we demonstrate that the longstanding discrepancy between theory and experiment does not originate from neglected DNA-specific elastic effects, but is explained by the indirect transmission of force through the handles, establishing it as the dominant missing ingredient in the theory of DNA looping under force.
\end{abstract}

\maketitle

\paragraph*{Introduction.}

Single-molecule force spectroscopy has become an indispensable tool for 
probing the mechanics and dynamics of biomolecules at the 
nanoscale~\cite{halma2024life, fazal2011optical}. Optical and magnetic 
tweezers, in particular, have enabled direct measurements of 
force-extension relations~\cite{smith1992direct}, torsional 
stiffness~\cite{kriegel2018measuring}, and looping and unfolding 
kinetics~\cite{chen2010protein, wen2007force, manosas2007force} for nucleic acids and 
proteins. A unifying theoretical framework for interpreting these 
experiments is the effective free energy landscape: by projecting the 
high-dimensional conformational dynamics of a polymer onto a single 
reaction coordinate, typically the end-to-end distance $r$, one obtains 
a one dimensional diffusion problem whose barrier height directly 
determines the timescale of the process of 
interest~\cite{woodside2014reconstructing, szabo1980first, shin2012effects, 
laeremans2025theoretical}.

\begin{figure}[t]
    \centering
    \includegraphics[width=\linewidth]{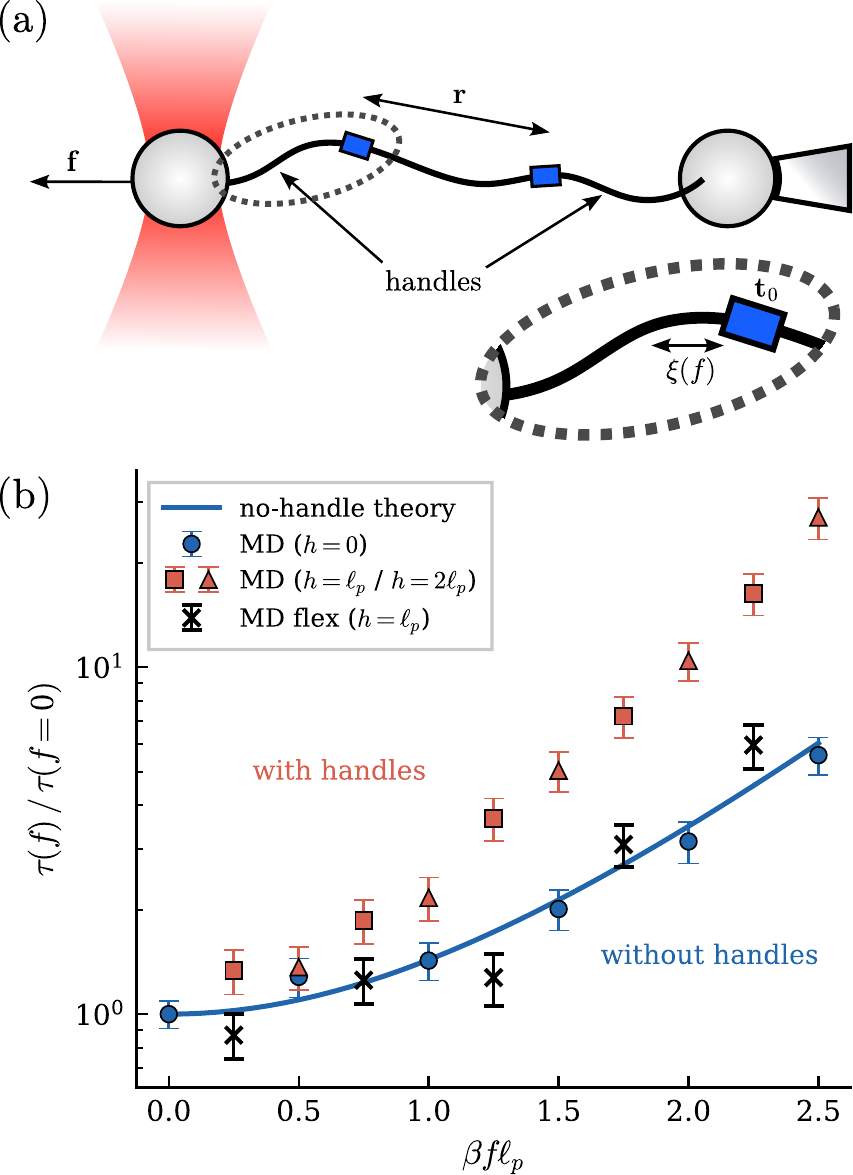}
    \caption{(a) Schematic of a DNA looping experiment in an optical-tweezer setup. The force $\mathbf{f}$ acts through DNA handles connecting the binding sites (blue squares) to the trapped beads. The interior segment, with end-to-end distance $r$, forms the loop. Inset: a handle leaves its binding site along the end tangent $\mathbf t_0$ of the interior segment and keeps this direction over the memory length $\xi(f)$.
(b)~Normalised looping time $\tau(f)/\tau(0)$ versus $\beta f\ell_p$ from MD simulations without handles ($h=0$, blue circles), with handles of length $h=\ell_p$ and $2\ell_p$ (orange squares/triangles), and with a flexible junction ($h=\ell_p$, black crosses). Solid line: no-handle theory. Handles make the looping time far more force sensitive, whereas a flexible junction recovers the no-handle result. Parameters correspond to $L\approx105\,\mathrm{nm}$, $\ell_p\approx50\,\mathrm{nm}$ and forces of order $10^2\,\mathrm{fN}$, representative of the experiments of Refs.~\cite{chen2010femtonewton,chen2010protein} (see SM).}
    \label{fig:setup_looping}
\end{figure}

A particularly important application is DNA looping, which plays a 
central role in gene regulation~\cite{milstein2011role}. Loop formation 
brings two specific binding sites into contact, and its kinetics are 
sensitive to the mechanical tension acting on the DNA~\cite{blumberg2005disruption, chen2010femtonewton, chen2010protein, milstein2011role}. Such tension arises 
both in single-molecule experiments and in biological contexts, where DNA 
is subject to forces from, for example, molecular motors, supercoiling, or 
chromatin organisation~\cite{schleif1992dna, allemand2006loops, milstein2011role, matthews1992dna, chen2010femtonewton, gallet2009power}. The relevance of this question is reinforced by a growing body of work proposing that chromatin loops themselves may form through the passive diffusive capture of anchor sites, rather than through active, motor-driven extrusion~\cite{gerguri2021comparison, uhlmann2025unified}, which would place the diffusion-limited looping kinetics considered here directly at the heart of chromosome organisation. Crucially, in all settings the tension acts 
globally on the DNA rather than specifically at the looping sites: the 
flanking DNA segments transmit the force to the looping region 
indirectly. In optical-tweezer experiments this is made explicit by the 
use of DNA handles that connect the binding sites to the trapped beads, 
as illustrated in Fig.~\ref{fig:setup_looping}a. The standard theoretical framework for looping kinetics under force assumes that the tension acts directly at the ends of the looping segment~\cite{blumberg2005disruption, shin2012effects, laeremans2024polymer, laeremans2025theoretical}, which is never strictly the case in practice. Consequently, discrepancies between theory and experiment have typically been attributed to shortcomings of the underlying polymer model, such as neglected DNA-specific elastic effects like twist~\cite{laeremans2025theoretical}. How indirect force transmission through the flanking DNA segments themselves affects the looping kinetics has not been systematically investigated. 

Here we ask, and answer, a precise question: how do flanking DNA segments influence the force dependence of DNA looping, and how can this effect be captured theoretically? Using molecular dynamics simulations, we find that handles of length comparable to the persistence length cause the looping time to increase substantially more steeply with force than predicted by standard theories that assume direct force application. The effect saturates once the handle length exceeds the persistence length and disappears entirely when the junction is made flexible, identifying orientational coupling as its physical origin. We explain these observations as a modification of the effective free energy landscape of the looping segment, which arises from indirect force transmission through the handles and can be incorporated directly into standard kinetic theories of looping.

This modification is set by two quantities. The first is the alignment of the end tangents of the looping segment with its end-to-end vector, extracted from simulations without applied force. The second is the length over which a handle remembers the orientation of its binding site. This memory length equals the persistence length at small force and shortens at larger force. With both quantities known, the force dependence of the looping kinetics follows analytically, without any finite-force simulations. The resulting theory agrees quantitatively with the experimental measurements of Chen~\textit{et al.}~\cite{chen2010femtonewton,chen2010protein}, without adjustable parameters. The indirect transmission of force through the handles thus explains the previously unresolved discrepancy between theory and experiment.

\paragraph*{Handle effect: simulation evidence.}

We model DNA as a wormlike chain (WLC) simulated by molecular dynamics (MD); full details are provided in the Supplemental Material (SM). The looping time $\tau(f)$ is the mean first-passage time for the end-to-end distance $r$ of the interior segment to reach a capture radius $r_\mathrm{c}$, the threshold separation at which the two binding sites are considered bound, starting from equilibrium. We compare three cases: (i)~no handles ($h=0$), (ii)~WLC handles of length $h=\ell_p$ and $h=2\ell_p$, where $\ell_p$ is the persistence length, and (iii)~handles connected to the interior segment through flexible junctions of zero bending rigidity.

Figure~\ref{fig:setup_looping}b shows $\tau(f)/\tau(0)$ as a function of the dimensionless force $\beta f \ell_p$. For a WLC under force $f$ without handles, the end-to-end probability distribution is~\cite{laeremans2025theoretical} \begin{equation} 
    P(r;f) \propto P_0(r)\, \frac{\sinh(\beta f r)}{\beta f r}. \label{eq:nohandle}
\end{equation} 
A derivation is provided in the SM. The looping time is then predicted from the effective free energy $\beta F(r;f)=-\ln P(r;f)$ using the Szabo--Schulten--Schulten theory~\cite{szabo1980first,shin2012effects,laeremans2025theoretical}, which models the dynamics of the end-to-end distance $r(t)$ as overdamped diffusion along the one-dimensional free energy landscape $F(r;f)$,
\begin{equation}
    \tau(f) \propto \frac{1}{Z} \int_{r_\mathrm{c}}^{L}\!dr \int_{r_\mathrm{c}}^{r}\!dr' \int_{r'}^{L}\!dr''\, \frac{ e^{-\beta F(r;f)} e^{-\beta F(r'';f)} }{ e^{-\beta F(r';f)} }, \label{eq:Szabo}
\end{equation} 
with $Z=\int_{r_\mathrm{c}}^{L} e^{-\beta F(r;f)}\,dr$. Ref.~\cite{laeremans2025theoretical} showed that this expression is accurate for short, stiff chains ($L/\ell_p\lesssim3$), precisely the regime relevant to the DNA looping experiments considered here~\cite{chen2010protein, chen2010femtonewton}. Consistent with this prediction, the solid blue curve accurately reproduces the no-handle simulations (blue circles) in Fig.~\ref{fig:setup_looping}b. Here $P_0(r)$ in Eq.~\eqref{eq:nohandle} can be obtained directly from simulation or analytically from mean-field approximations (see SM). With handles, however, the looping time increases much more steeply with force than predicted by the no-handle theory. Strikingly, the results for $h=\ell_p$ (orange squares) and $h=2\ell_p$ (orange triangles) are virtually identical, demonstrating that the effect saturates once $h\gtrsim\ell_p$, a feature to which we return in the next section. The third simulation, with the handles joined flexibly to the looping region, serves as a control experiment: Here (black crosses), the enhancement disappears entirely, confirming that the orientational stiffness at the attachment point is a crucial element of the described effect.

\begin{figure*}[t]
    \centering
    \includegraphics[width=\linewidth]{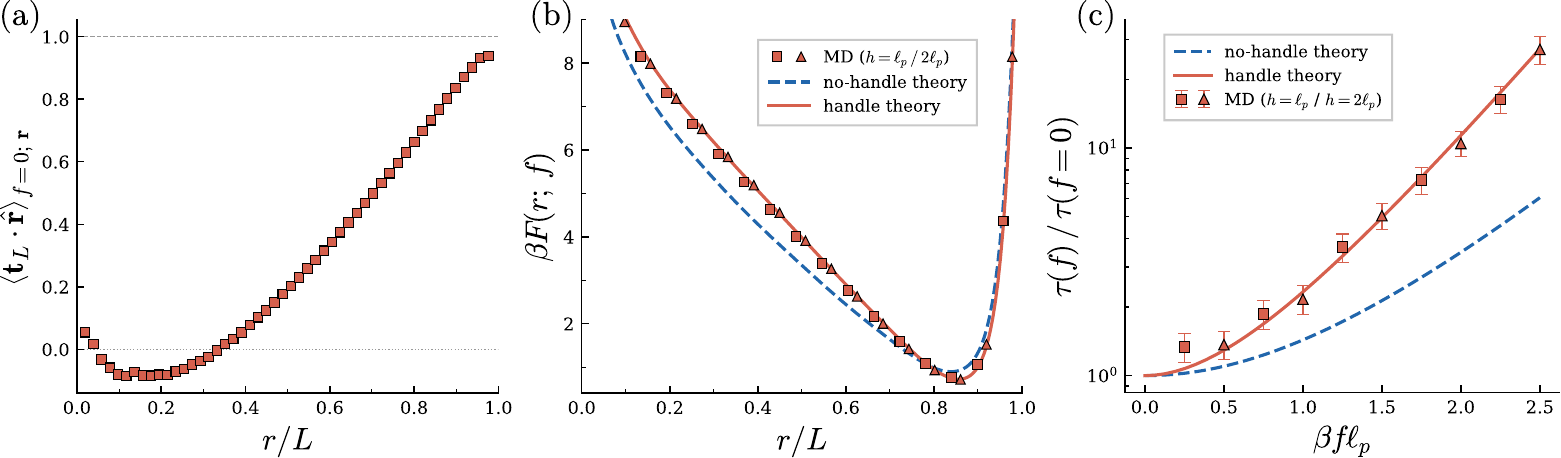}
    \caption{Validation of the handle theory. (a) End-tangent alignment $\langle\mathbf t_L\cdot\hat{\mathbf r}\rangle_{f=0;\,\mathbf r}$ of the interior segment, extracted from zero-force MD simulations with $h=\ell_p$ (see SM).
    (b) Effective free energy $\beta F(r;f)$ at $\beta f \ell_p = 1.5$: MD data for
    $h=\ell_p$ (orange squares) and $h=2\ell_p$ (orange triangles) versus
    the handle theory, Eqs.~\eqref{eq:handle} and~\eqref{eq:xi}, with the alignment of panel~(a)
    (orange line) and the no-handle theory (blue dashed).
    (c) Normalised looping time $\tau(f)/\tau(0)$: MD simulations with
    handles (orange symbols), no-handle theory (blue dashed), and handle theory, Eqs.~\eqref{eq:handle} and~\eqref{eq:xi}
    (orange solid).}
    \label{fig:theory}
\end{figure*}

\paragraph*{Analytic framework with handles.}

When handles are present, the force acts on the full construct of total length $L_\mathrm{tot}=2h+L$ rather than directly on the interior segment. Starting from the exact path integral expression for the interior segment distribution (see SM), we obtain 
\begin{equation} 
    P(\mathbf r;\mathbf f)= \frac{ P_0(\mathbf r)\, e^{\beta\mathbf f\cdot\mathbf r}\, C[\mathbf r;\mathbf f] } { \int d^3r'\, P_0(\mathbf r')\, e^{\beta\mathbf f\cdot\mathbf r'}\, C[\mathbf r';\mathbf f] }, \label{eq:segment_final} 
\end{equation} 
where 
\begin{equation} 
    C[\mathbf r;\mathbf f] = \left\langle e^{\beta\mathbf f\cdot\mathbf R_\mathrm{handles}} \right\rangle_{f=0;\mathbf r}
\end{equation} 
is a conditional average over the handle degrees of freedom, with $\mathbf R_\mathrm{handles}$ defined through $\mathbf R_\mathrm{tot}=\mathbf r+\mathbf R_\mathrm{handles}$, where $\mathbf R_\mathrm{tot}$ is the end-to-end vector of the full construct. The factor $C[\mathbf r;\mathbf f]$ captures the orientational correlations between the handles and the interior segment, for $h=0$ one has $C=1$ and Eq.~\eqref{eq:nohandle} is recovered exactly. 

The handle effect can be understood from a simple picture. Without handles, Eq.~\eqref{eq:nohandle} follows from reweighting the zero-force distribution by the work $fr\cos\theta$ that the force does on the end-to-end vector, averaged over all orientations $\theta$ of $\mathbf r$ relative to the force (see SM). With handles, the force does not act on the ends of the looping segment but on the far ends of the handles. Because the WLC penalises bending, each handle leaves its binding site along the unit end tangent of the looping segment, $\mathbf t_0$ or $\mathbf t_L$, and keeps this direction over a finite length $\xi(f)$ before bending fluctuations and the force erase it (Fig.~\ref{fig:setup_looping}a). Each end of the looping segment thus carries an effective rigid extension of length $\xi$ along its end tangent, and the force pulls on $\mathbf r+\xi(\mathbf t_0+\mathbf t_L)$ rather than on $\mathbf r$. At fixed $\mathbf r$, the end tangents fluctuate, but on average only their components along $\hat{\mathbf r}=\mathbf r/r$ survive, the perpendicular components cancel by symmetry. Replacing the end tangents by this average is exact to first order in $f$ at small force (see SM). The force therefore effectively pulls on a vector of length $r+\Delta$ along $\hat{\mathbf r}$, and the orientational average proceeds exactly as without handles (see SM):
\begin{equation} 
    \begin{gathered}
    P(r;f) \propto P_0(r)\, \frac{\sinh\!\bigl[\beta f(r+\Delta)\bigr]} {\beta f(r+\Delta)},\\
\Delta(r;f)=2\,\xi(f)\,\langle\mathbf t_L\cdot\hat{\mathbf r}\rangle_{f=0;\mathbf r}.
    \end{gathered} \label{eq:handle} 
\end{equation}
Compared with Eq.~\eqref{eq:nohandle}, the sole effect of the handles is the replacement $r\to r+\Delta$. The shift combines a property of the handles, the memory length $\xi(f)$, with a zero-force property of the looping segment, the alignment $\langle\mathbf t_L\cdot\hat{\mathbf r}\rangle_{f=0;\mathbf r}$ of its end tangents with its end-to-end vector.

The end-tangent alignment $\langle\mathbf t_L\cdot\hat{\mathbf r}\rangle_{f=0;\mathbf r}$ can be extracted directly from a zero-force simulation (see SM). The force-dependent memory length $\xi(f)$ can be estimated from theory. At small force, we simply have $\xi_\text{small}(f)=\ell_p$, since tangent correlations decay as $\langle\mathbf t(s)\!\cdot\!\mathbf t(s')\rangle=e^{-|s-s'|/\ell_p}$. At large force the handle is pulled back into line with the force more quickly, and $\xi$ reduces to the deflection length $\xi_\text{large}(f) = \sqrt{\ell_p/\beta f}$~\cite{odijk1995stiff,marantan2018mechanics}. To bridge both limits, we note that a weak force only slightly reweights the chain conformations, so that $\xi$ must decrease linearly with $f$ at small force. The simplest interpolation that preserves this is to add their inverse squares, $1/\xi^2=1/\xi_\text{small}^2+1/\xi_\text{large}^2=1/\ell_p^2+\beta f/\ell_p$, which recovers both limits:
\begin{equation}
    \xi(f) \approx \frac{\ell_p}{\sqrt{1+\beta f\ell_p}}. \label{eq:xi}
\end{equation}
This assumes handles longer than $\xi$, as is the case in experiments. Eq.~\eqref{eq:xi} immediately explains the observations of Fig.~\ref{fig:setup_looping}b. Since $\xi\le\ell_p$, handle segments further than a persistence length from the binding site no longer transmit its orientation, so the effect saturates once $h\gtrsim\ell_p$. A flexible junction removes the bending penalty at the binding site, so that the handles no longer follow $\mathbf t_0$ and $\mathbf t_L$, and Eq.~\eqref{eq:nohandle} is recovered.

Figure~\ref{fig:theory} validates the handle theory against MD simulations (sampling protocol in SM). Panel~(a) shows the end-tangent alignment $\langle\mathbf t_L\cdot\hat{\mathbf r}\rangle_{f=0;\mathbf r}$, which, together with $\xi(f)$, sets the shift $\Delta$ in Eq.~\eqref{eq:handle}. It approaches unity as $r\to L$, where the interior segment is nearly straight and its end tangents point along $\mathbf r$, and becomes small, even slightly negative, for $r\lesssim0.3L$, where the segment bends back on itself. Handles therefore lengthen the looping coordinate most strongly at large $r$. Panel~(b) compares the resulting free energy $\beta F(r;f)=-\ln P(r;f)$, obtained from Eqs.~\eqref{eq:handle} and~\eqref{eq:xi}, with direct MD measurements at $\beta f \ell_p = 1.5$. Relative to the no-handle prediction (blue dashed line), the presence of handles produces a substantially steeper free energy landscape, implying a stronger suppression of loop formation by force. The theoretical curve (orange solid line) accurately reproduces the simulation data for both $h=\ell_p$ and $h=2\ell_p$, confirming again that the handle effect saturates once the handle length exceeds a persistence length. Panel~(c) shows the corresponding normalized looping times. The handle theory (orange line) is in good agreement with the MD results over the full force range considered, whereas the no-handle theory (blue dashed line) significantly underestimates the force induced increase in looping time. This demonstrates that the enhanced force sensitivity arises from the orientational coupling transmitted through the handles and is quantitatively captured by the shift $\Delta(r;f)$, built from a zero-force property of the looping segment and the memory length of the handles.

\paragraph*{Comparison with experiment.}

\begin{figure}[t]
    \centering
    \includegraphics[width=\linewidth]{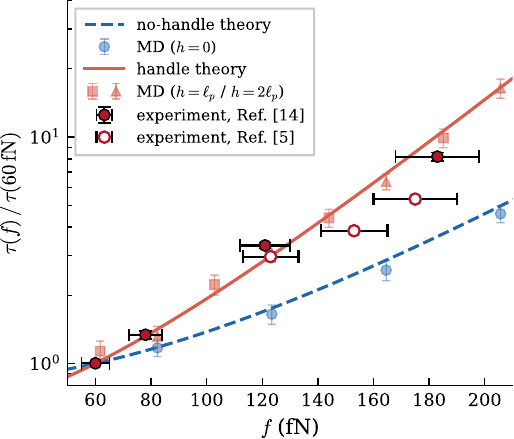}
    \caption{Force-dependent looping time, normalized by its value at $f=60\,\mathrm{fN}$, compared with constant-force measurements of Chen \textit{et al.}: filled red circles from Ref.~\cite{chen2010femtonewton}, open red circles from the zero-noise measurements of Ref.~\cite{chen2010protein}. Blue dashed line: no-handle theory. Semi-transparent blue circles: no-handle MD simulations. Orange solid line: handle theory, Eqs.~\eqref{eq:handle} and~\eqref{eq:xi}. Semi-transparent orange squares and triangles: MD simulations with handles of length $h=\ell_p$ and $h=2\ell_p$, respectively. Simulations are matched to the corresponding theory at zero force, and theory and simulations are then normalized by the theory at $60\,\mathrm{fN}$; all experimental data are normalized by the measurement at $60\,\mathrm{fN}$ of Ref.~\cite{chen2010femtonewton} (see SM). Vertical error bars follow from the reported uncertainties of the looping rates; none are reported for the open circles.}
    \label{fig:experiment}
\end{figure}

Having established the handle effect in simulations, we now test whether it is sufficient to explain the experimental observations of Chen~\textit{et al.}~\cite{chen2010femtonewton,chen2010protein}. Figure~\ref{fig:experiment} compares the measured force dependence of the looping time with direct MD simulations of the wormlike chain model. The experimental data comprise two sets of constant-force measurements on the same construct: a dedicated study of loop formation under constant tension~\cite{chen2010femtonewton} (filled circles), and zero-noise reference measurements from a subsequent study of fluctuating forces~\cite{chen2010protein} (open circles). The experiments employed a 1316 bp DNA construct containing a 305 bp looping segment, corresponding to an interior contour length of approximately $103.7\,\mathrm{nm}$. Using the mapping described in the SM, the simulated wormlike chain has an interior contour length of around $105\,\mathrm{nm}$, closely matching the experimental geometry. Remarkably, the simulations with handles (semi-transparent orange symbols) reproduce the experimental force dependence, in particular that of the dedicated constant-force study over its full force range, whereas simulations without handles (semi-transparent blue circles) substantially underestimate the increase in looping time with force. This is particularly significant because DNA-specific effects, such as twist and binding-site phasing, had previously been proposed as possible explanations for the failure of no-handle theories at short DNA lengths~\cite{laeremans2025theoretical}. The agreement between experiment and the simple handled wormlike chain model shows that indirect force transmission through the handles is the dominant factor responsible for the previously unresolved discrepancy between theory and experiment.

We next test the analytic theory. As shown above, the force-dependent looping kinetics can be predicted from two zero-force quantities of the looping segment, $P_0(r)$ and $\langle\mathbf t_L\cdot\hat{\mathbf r}\rangle_{f=0;\mathbf r}$, both obtained from a single zero-force simulation, together with the memory length $\xi(f)$ of Eq.~\eqref{eq:xi}. Using these as input to Eq.~\eqref{eq:handle}, we obtain the orange solid curve in Fig.~\ref{fig:experiment}. The theory agrees with the simulations with handles over the full force range, and with the constant-force measurements of Ref.~\cite{chen2010femtonewton} at all forces from $60$ to $183\,\mathrm{fN}$, whereas the no-handle theory underestimates the looping time at $183\,\mathrm{fN}$ by more than a factor of two. The zero-noise measurements of Ref.~\cite{chen2010protein} agree at $123\,\mathrm{fN}$, but lie below the theory and the simulations at $153$ and $175\,\mathrm{fN}$. This deviation is not caused by the approximations of the theory, since theory and simulations agree with each other. Instead, it is of the order of the difference between the two experiments: at nearly equal forces ($121$ and $123\,\mathrm{fN}$), the two data sets differ by approximately $10\%$. Moreover, the second data set lacks its own measurement at $60\,\mathrm{fN}$ and is normalized using the first, so that a systematic offset between the experiments shifts all its points by the same factor. The theory thus has predictive power for DNA looping experiments, quantitatively reproducing the measured force dependence of the looping kinetics without adjustable parameters and without any simulation at finite force.

\paragraph*{Discussion and conclusion.}

We have shown that the geometry of force transmission plays a decisive role in DNA looping kinetics. While previous theories implicitly assumed that force acts directly on the looping segment, most single-molecule experiments apply force through flanking DNA handles. We find that these handles amplify the force dependence of looping, increasing the looping time relative to the standard no-handle prediction. The effect emerges from orientational correlations between the handles and the looping segment and saturates once the handle length exceeds approximately one persistence length, reflecting the finite range of tangent--tangent correlations along DNA. Flexible junctions suppress the effect entirely, confirming that it is transmitted through bending rigidity at the attachment point. 


We further show that the handle effect observed in our simulations is sufficient to explain the experimental measurements of Chen \textit{et al.}~\cite{chen2010femtonewton,chen2010protein}. Simulations that include handles reproduce the experimentally observed force dependence of the looping time, in particular over the full force range of the dedicated constant-force study~\cite{chen2010femtonewton}, whereas simulations without handles substantially underestimate the effect. This is noteworthy because previous work had suggested that the disagreement between experiment and no-handle theories might originate from missing DNA-specific physics~\cite{laeremans2025theoretical}. Instead, our results show that a simple wormlike chain model is already sufficient once indirect force transmission through the handles is taken into account, identifying handle-induced orientational coupling as the dominant missing ingredient.

We then show that the same effect can be predicted without simulations at finite force. Eq.~\eqref{eq:handle} reduces the handle effect to a shift of the looping coordinate, set by the zero-force alignment between the end tangents and the end-to-end vector of the looping segment $\langle\mathbf t_L\cdot\hat{\mathbf r}\rangle_{f=0;\mathbf r}$, and by the memory length $\xi(f)$ over which a handle keeps the orientation of its binding site. Combined with the standard Szabo--Schulten--Schulten framework, this yields a parameter-free prediction for the force dependence of looping kinetics. The resulting theory quantitatively reproduces our simulations and the constant-force measurements over the full force range, with remaining deviations for part of the data of Ref.~\cite{chen2010protein} that are of the order of the differences between the two experiments.

More broadly, our work shows that force-transmitting elements can play an active role in determining molecular kinetics. In DNA looping, the handles do not merely relay the applied force to the looping segment, they modify how the force is transmitted through orientational coupling. This effect is sufficiently strong to explain a longstanding discrepancy between theory and experiment and cannot be accounted for by refinements of the underlying DNA elasticity alone. Similar effects may arise whenever a reactive coordinate is coupled to an externally loaded structure, including nucleic acid folding experiments performed with optical tweezers~\cite{wen2007force,manosas2007force} and potentially the passive diffusive encounter of chromatin loci in vivo~\cite{gerguri2021comparison,uhlmann2025unified}. By expressing the correction entirely in terms of equilibrium observables, our framework provides a practical route to incorporate such effects into existing theoretical descriptions.


\paragraph*{Acknowledgements ---}
This research is financially supported by the Dutch Ministry of Education,
Culture and Science (Gravity Program 024.005.020 – Interactive Polymer
Materials IPM).

\paragraph*{Data availability ---}
Code and data belonging to the findings of the work will be made publicly available upon acceptance.

\bibliography{references}


\clearpage
\onecolumngrid 
\setcounter{section}{0} \renewcommand{\thesection}{S\arabic{section}} \renewcommand{\theequation}{S\arabic{equation}} 
\renewcommand{\thefigure}{S\arabic{figure}} \setcounter{equation}{0} 
\setcounter{figure}{0} 
\setcounter{page}{1}

\begin{center} 
    {\Large\bfseries Supplemental Material}\\[0.5em] {\large DNA handles bias force-dependent looping times}\\[1em] Wout Laeremans$^{1}$, Jef Hooyberghs$^{2}$, and Wouter G. Ellenbroek$^{1}$\\[0.5em] $^{1}$Soft Matter and Biological Physics, Department of Applied Physics and Science Education, and Institute for Complex Molecular Systems, Eindhoven University of Technology, P.O. Box 513, 5600 MB Eindhoven, The Netherlands\\ $^{2}$UHasselt, Faculty of Sciences, Data Science Institute, Theory Lab, Agoralaan, 3590 Diepenbeek, Belgium 
\end{center} 

\vspace{1em}

This Supplemental Material provides (I) a review of the standard WLC
theory for the effective free energy and looping time without handles,
(II) a complete derivation of the handle correction factor and the
modified distribution, Eq.~(5) of the main text, and (III) simulation details, including the extraction of the end-tangent alignment and the normalization of the looping times.
 
\section{I. WLC free energy and looping time without handles}
\label{sec:SM_nohandles}

We model the polymer as a wormlike chain (WLC)~\cite{kratky1949rontgenuntersuchung, marko1995stretching}, the standard continuum description of semiflexible biopolymers such as double stranded DNA. The polymer has contour length $L$ and is characterised by a unit tangent field $\mathbf{t}(s)$, with curvilinear distance $s \in [0, L]$. In the absence of an applied force, the Hamiltonian reads
\begin{equation}
    \beta \mathcal{H}_0[\mathbf{t}] = \frac{\ell_p}{2} \int_0^{L} \left|\partial_s \mathbf{t}(s)\right|^2 ds,
\end{equation}
where $\beta^{-1} = k_\mathrm{B}T$ is the thermal energy and $\ell_p$ the persistence length. The end-to-end vector is
\begin{equation}
    \mathbf{r}[\mathbf{t}] = \int_0^L \mathbf{t}(s)\, ds.
\end{equation}
When a constant force $\mathbf{f}$ is applied, as in an optical-tweezer experiment, the Hamiltonian acquires an additional mechanical work term,
\begin{equation}
    \beta \mathcal{H}_f[\mathbf{t}] = \beta \mathcal{H}_0[\mathbf{t}] - \beta \mathbf{f} \cdot \mathbf{r}[\mathbf{t}].
\end{equation}

\subsection{Probability density under force}
The probability density of observing a given end-to-end vector $\mathbf{r}$ at force $\mathbf{f}$ is obtained from the path integral
\begin{equation}
    P(\mathbf{r};\mathbf{f}) = \frac{\displaystyle \int \mathcal{D}[\mathbf{t}]\; \delta\!\left(\mathbf{r} - \mathbf{r}[\mathbf{t}]\right) e^{-\beta \mathcal{H}_f[\mathbf{t}]}}{\displaystyle \int \mathcal{D}[\mathbf{t}]\; e^{-\beta \mathcal{H}_f[\mathbf{t}]}}.
\end{equation}
Denoting the zero force distribution as $P_0(\mathbf{r}) \equiv P(\mathbf{r};\mathbf{f}=0)$, the distribution under force can be written in closed form as~\cite{laeremans2025theoretical}
\begin{equation}
    P(\mathbf{r};\mathbf{f}) = \frac{P_0(\mathbf{r})\, e^{\beta \mathbf{f} \cdot \mathbf{r}}}{\displaystyle \int d^3r'\, P_0(\mathbf{r}')\, e^{\beta \mathbf{f} \cdot \mathbf{r}'}}.
    \label{eq:Pnohandles}
\end{equation}
Without loss of generality we orient the force along the $z$-axis, $\mathbf{f} = f\hat{\mathbf{z}}$. At zero force the chain is isotropic, so $P_0(\mathbf{r}) \sim P_0(r)$ depends only on the end-to-end distance $r = |\mathbf{r}|$. Integrating Eq.~\eqref{eq:Pnohandles} over the solid angle using $\mathbf{f}\cdot\mathbf{r} = fr\cos\theta$ gives~\cite{laeremans2025theoretical}
\begin{equation}
    P(r;f) \propto P_0(r) \int_0^{\pi} \sin\theta\, e^{\beta f r \cos\theta}\, d\theta
             = P_0(r)\,\frac{\sinh(\beta f r)}{\beta f r}, \label{eq:calcangle}
\end{equation}
as was given in the main text in Eq.~\eqref{eq:nohandle}. The corresponding effective free energy $\beta F(r;f) = -\ln P(r;f)$ then satisfies~\cite{laeremans2025theoretical}
\begin{align}
    \beta F(r;f) = \beta F(r;0) - \ln\!\left[\frac{\sinh(\beta f r)}{\beta f r}\right].
    \label{eq:WLtheory}
\end{align}

\begin{figure}[t]
    \centering
    \includegraphics[width=0.9\linewidth]{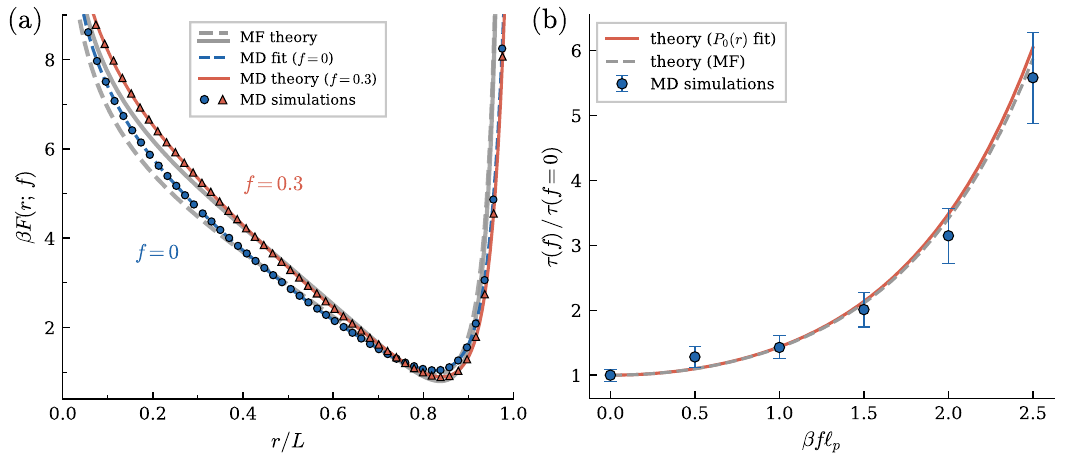}
    \caption{(a) Effective free energy $\beta F(r;f)$ of a WLC with $L/\ell_p=2.1$. Blue circles: zero force MD data. Blue dashed line: 12th-order polynomial fit used to represent $P_0(r)$. Orange triangles: MD data at $f=0.3$ (simulation units, $\beta f\ell_p=1.5$). Orange solid line: prediction from $\beta F(r;f)=\beta F(r;0)-\ln[\sinh(\beta fr)/(\beta fr)]$ using the fitted zero force free energy as input. Grey dashed and solid lines: corresponding predictions obtained using the mean-field approximation, Eq.~\eqref{eq:MF}, for $P_0(r)$. (b) Normalized looping time $\tau(f)/\tau(0)$ obtained from the Szabo--Schulten--Schulten theory, Eq.~\eqref{eq:Szabo}. Blue line: prediction using the simulation-derived $P_0(r)$. Grey line: prediction using the mean-field approximation, Eq.~\eqref{eq:MF}. Symbols denote MD simulations. Both choices of $P_0(r)$ provide accurate predictions for the force-dependent looping time.}
    \label{fig:SUP1_PRL}
\end{figure}

\subsection{Simulation versus analytical approximations for $P_0(r)$} Equation~\eqref{eq:WLtheory} reduces the finite force problem to the zero force free energy $\beta F(r;0) = -\ln P_0(r)$. In the main text, $P_0(r)$ was obtained directly from MD simulations using the procedure of Ref.~\cite{laeremans2025theoretical}. That is, by sampling end-to-end distances in the MD simulation, binning them, and building a normalized probability density. The resulting zero force free energy, $\beta F(r;0)=-\ln P_0(r)$, is shown in Fig.~\ref{fig:SUP1_PRL}(a) as blue circles and fitted with a 12th-order polynomial (blue dashed line). The corresponding simulation data at $f=0.3$ (orange triangles, force reported in simulation units, $\beta f\ell_p=1.5$) are accurately reproduced by
\begin{equation} 
    \beta F(r;f) = \beta F(r;0) - \ln\!\left[ \frac{\sinh(\beta f r)} {\beta f r} \right], 
\end{equation} 
using the fitted zero force free energy as input (orange solid line). Although no exact closed-form expression is known for $P_0(r)$, several analytical approximations have been proposed \cite{wilhelm1996radial,bhattacharjee1997distribution}. One example is the mean-field (MF) approximation of Bhattacharjee \textit{et al.}~\cite{bhattacharjee1997distribution}, 
\begin{equation} 
    P_0^{\mathrm{MF}}(r) \propto r^{2} \left[1-\left(\frac{r}{L}\right)^2\right]^{-9/2} \exp\!\left[ -\frac{3L} {4\ell_p\left(1-\left(\frac{r}{L}\right)^2\right)} \right], \label{eq:MF} 
\end{equation} 
which reproduces both the Gaussian limit at small extensions and the correct divergence as $r\to L$. The resulting free energies are shown as grey lines in Fig.~\ref{fig:SUP1_PRL}(a). While less accurate than the simulation-derived free energy, the MF approximation captures the overall shape well. Both forms of $P_0(r)$ can be inserted into the Szabo--Schulten--Schulten expression, Eq.~\eqref{eq:Szabo}, to predict the looping time. The resulting curves are shown in Fig.~\ref{fig:SUP1_PRL}(b). Despite the small differences in the free energy landscapes, the predicted looping times are nearly indistinguishable and both agree with the simulation data.

\section{II. Derivation of the handle correction}
\label{sec:SM_derivation2}

We label the arc-length coordinate of the full polymer by $s\in[0,L_{\mathrm{tot}}]$ and denote the interior looping segment by $s\in[s_0,s_0+L]$. Its end-to-end vector is 
\begin{equation} 
    \mathbf r[\mathbf t] = \int_{s_0}^{s_0+L}\mathbf t(s)\,ds, 
\end{equation} 
while the total end-to-end vector decomposes as 
\begin{equation} 
    \mathbf R_{\mathrm{tot}}[\mathbf t] = \mathbf r[\mathbf t] + \mathbf R_{\mathrm{handles}}[\mathbf t], 
\end{equation} 
where 
\begin{equation} 
    \mathbf R_{\mathrm{handles}}[\mathbf t] = \int_0^{s_0}\mathbf t(s)\,ds + \int_{s_0+L}^{L_{\mathrm{tot}}}\mathbf t(s)\,ds. 
\end{equation}
Because the force is applied to the ends of the full construct, the Boltzmann weight contains $e^{\beta\mathbf f\cdot\mathbf R_{\mathrm{tot}}}$. The probability density for the end-to-end vector of the interior segment is therefore 
\begin{equation} 
    P(\mathbf r;\mathbf f)= \frac{ \displaystyle\int\mathcal D[\mathbf t]\, \delta\!\left(\mathbf r-\mathbf r[\mathbf t]\right) e^{-\beta\mathcal H_0[\mathbf t] +\beta\mathbf f\cdot\mathbf R_{\mathrm{tot}}[\mathbf t]} } { \displaystyle\int\mathcal D[\mathbf t]\, e^{-\beta\mathcal H_0[\mathbf t] +\beta\mathbf f\cdot\mathbf R_{\mathrm{tot}}[\mathbf t]} }, \label{eq:path_split_SM} 
\end{equation} 
where $\mathcal H_0$ is the zero force WLC Hamiltonian of the full construct. Using the decomposition $\mathbf R_{\mathrm{tot}} = \mathbf r+\mathbf R_{\mathrm{handles}}$, the numerator of Eq.~\eqref{eq:path_split_SM} factorises as 
\begin{align} 
    & \int\mathcal D[\mathbf t]\, \delta\!\left(\mathbf r-\mathbf r[\mathbf t]\right) e^{-\beta\mathcal H_0[\mathbf t] +\beta\mathbf f\cdot\mathbf R_{\mathrm{tot}}[\mathbf t]} = e^{\beta\mathbf f\cdot\mathbf r} \int\mathcal D[\mathbf t]\, \delta\!\left(\mathbf r-\mathbf r[\mathbf t]\right) e^{-\beta\mathcal H_0[\mathbf t] +\beta\mathbf f\cdot \mathbf R_{\mathrm{handles}}[\mathbf t]}, 
\end{align} 
where the factor $e^{\beta\mathbf f\cdot\mathbf r}$ has been pulled outside the integral using the delta function constraint $\mathbf r[\mathbf t]=\mathbf r$. Multiplying and dividing by the zero force distribution 
\begin{equation} 
    P_0(\mathbf r) = \frac{ \displaystyle\int\mathcal D[\mathbf t]\, \delta\!\left(\mathbf r-\mathbf r[\mathbf t]\right) e^{-\beta\mathcal H_0[\mathbf t]} } { \displaystyle\int\mathcal D[\mathbf t]\, e^{-\beta\mathcal H_0[\mathbf t]} }, 
\end{equation} 
the numerator becomes proportional to 
\begin{equation} 
    e^{\beta\mathbf f\cdot\mathbf r}\, P_0(\mathbf r)\, C[\mathbf r;\mathbf f], 
\end{equation} 
where we define the handle correction factor 
\begin{align} 
    C[\mathbf r;\mathbf f] &= \frac{ \displaystyle\int\mathcal D[\mathbf t]\, \delta\!\left(\mathbf r-\mathbf r[\mathbf t]\right) e^{-\beta\mathcal H_0[\mathbf t] +\beta\mathbf f\cdot \mathbf R_{\mathrm{handles}}[\mathbf t]} } { \displaystyle\int\mathcal D[\mathbf t]\, \delta\!\left(\mathbf r-\mathbf r[\mathbf t]\right) e^{-\beta\mathcal H_0[\mathbf t]} } = \left\langle e^{\beta\mathbf f\cdot\mathbf R_{\mathrm{handles}}} \right\rangle_{f=0;\mathbf r}. \label{eq:C_def_SM} 
\end{align} 
This is the zero force conditional average of the handle Boltzmann weight, taken over all chain conformations with the interior segment end-to-end vector fixed at $\mathbf r$. Substituting this result into Eq.~\eqref{eq:path_split_SM} and performing the same decomposition in the denominator yields the result of Eq.~\eqref{eq:segment_final} of the main text:
\begin{equation} 
    P(\mathbf r;\mathbf f) = \frac{ P_0(\mathbf r)\, e^{\beta\mathbf f\cdot\mathbf r}\, C[\mathbf r;\mathbf f] } { \displaystyle \int d^3r'\, P_0(\mathbf r')\, e^{\beta\mathbf f\cdot\mathbf r'}\, C[\mathbf r';\mathbf f] }, \label{eq:segment_final_SM} 
\end{equation} 
which has the same structure as the no-handle result, with all effects of the handles encoded in the correction factor $C[\mathbf r;\mathbf f]$. When $h=0$, one has $C=1$ identically and the no-handle result is recovered exactly.

\subsection{Derivation of Eq.~\eqref{eq:handle}}
First, we write the handles more explicitly as $\mathbf R_\mathrm{tot}=\mathbf r+\mathbf R_1+\mathbf R_2$, with $\mathbf R_{1,2}$ the two handles. Then, we can write Eq.~\eqref{eq:segment_final_SM} as
\begin{equation}
P(\mathbf r;\mathbf f)\propto P_0(\mathbf r)\,e^{\beta\mathbf f\cdot\mathbf r}\,C(\mathbf r),\qquad
C(\mathbf r)=\big\langle e^{\beta\mathbf f\cdot(\mathbf R_1+\mathbf R_2)}\big\rangle_{f=0;\mathbf r}.
\end{equation}
The bending energy couples only neighbouring bonds, due to locality in the WLC model. Given the interior segment, each handle therefore depends only on its junction tangent, $\mathbf t_0$ or $\mathbf t_L$, so
\begin{equation}
C(\mathbf r)=\big\langle Q(\mathbf t_0)\,Q(\mathbf t_L)\big\rangle_{f=0;\mathbf r},\qquad
Q(\mathbf t)=\big\langle e^{\beta\mathbf f\cdot\mathbf R_h}\big\rangle_{\mathbf t},
\end{equation}
where $Q(\mathbf t)$ is the force weight of a single handle whose first bond points along $\mathbf t$, and $\mathbf R_h$ is the corresponding handle vector. Next, we treat each handle as a rigid lever of length $\xi(f)$ along its junction tangent, $\ln Q(\mathbf t)\simeq\mathrm{const}+\beta\xi(f)\,\mathbf f\cdot\mathbf t$. At small force this is exact to first order in $f$: $\ln Q\simeq\beta\mathbf f\cdot\langle\mathbf R_h\rangle_{\mathbf t}=\beta a_0\,\mathbf f\cdot\mathbf t$, with $a_0$ the mean extension of the handle along its junction tangent, Eq.~\eqref{eq:a0_SM}, which tends to $\ell_p$ for long handles. At larger force, $\xi(f)$ decreases as discussed in the main text. This gives
\begin{equation}
C(\mathbf r)\propto\big\langle e^{\beta\xi\,\mathbf f\cdot(\mathbf t_0+\mathbf t_L)}\big\rangle_{f=0;\mathbf r}.
\end{equation}
At fixed $\mathbf r$ and zero force, rotations about $\hat{\mathbf r}$ leave the energy unchanged, and reversing the chain swaps its two ends. Hence $\langle\mathbf t_0+\mathbf t_L\rangle_{f=0;\mathbf r}=2\langle\mathbf t_L\cdot\hat{\mathbf r}\rangle_{f=0;\mathbf r}\,\hat{\mathbf r}$. Replacing $\mathbf t_0+\mathbf t_L$ by this mean, gives
\begin{equation}
P(\mathbf r;\mathbf f)\propto P_0(\mathbf r)\,e^{\beta\mathbf f\cdot\hat{\mathbf r}\,(r+\Delta)},\qquad
\Delta=2\xi(f)\,\langle\mathbf t_L\cdot\hat{\mathbf r}\rangle_{f=0;\mathbf r}.
\end{equation}
Integrating over the direction of $\mathbf r$ as in Eq.~\eqref{eq:calcangle} yields Eq.~\eqref{eq:handle}. 

Replacing $\mathbf t_0+\mathbf t_L$ by its mean is the only approximation in this step, and we now discuss its validity. Writing $x=\beta\xi\,\mathbf f\cdot(\mathbf t_0+\mathbf t_L)$, the exact handle contribution to $\ln P$ is $\ln\langle e^{x}\rangle_{\mathbf r}$, whereas the approximation uses $\langle x\rangle_{\mathbf r}$. Expanding the exponential for small $x$,
\begin{equation}
\ln\langle e^{x}\rangle_{\mathbf r}=\langle x\rangle_{\mathbf r}+\tfrac12\mathrm{Var}_{\mathbf r}(x)+\dots,
\end{equation}
so both expressions agree up to first order in $x$ and differ only through the fluctuations of the end tangents at fixed $\mathbf r$, contained in $\mathrm{Var}_{\mathbf r}(x)$. The size of $x$ is set by
\begin{equation}
\beta f\xi=\frac{\beta f\ell_p}{\sqrt{1+\beta f\ell_p}}.
\end{equation}
At small force, $\xi\simeq\ell_p$, so that $x\propto f$. As for the lever approximation above, this step is then exact to first order in $f$, and the neglected fluctuations contribute only at order $f^2$. At larger force, $\xi$ decreases, but more slowly than $f$ increases, so that $\beta f\xi\sim\sqrt{\beta f\ell_p}$ keeps growing and the expansion no longer holds. Nevertheless the neglected terms become large only at forces where loop formation is already strongly suppressed.

\section{III. Simulation details} All molecular dynamics (MD) simulations were performed using LAMMPS~\cite{LAMMPS} in dimensionless Lennard--Jones units. The inverse temperature was fixed to $\beta=(k_\mathrm{B}T)^{-1}=1$. 

\subsection{Polymer model} 
The polymer was represented as a bead-spring chain of $N+1$ beads with equilibrium bond length $b=0.5$ and bead mass $m=1$. Consecutive beads were connected by harmonic bonds, 
\begin{equation} 
    U_\mathrm{bond} = \frac{K}{2} \sum_{i=0}^{N-1} \left( |\mathbf r_{i+1}-\mathbf r_i|-b \right)^2, 
\end{equation} 
with bond stiffness $K=10^4$, ensuring negligible bond length fluctuations. Bending rigidity was introduced through 
\begin{equation} 
    U_\mathrm{bend} = \frac{\kappa}{2} \sum_{i=1}^{N-1} \theta_i^2, 
\end{equation} 
where $\theta_i$ is the angle between consecutive bond vectors and $\kappa=10$. The corresponding persistence length is 
\begin{equation} 
    \ell_p=\beta\kappa b=5, 
\end{equation} 
where we used the standard wormlike chain relation $\ell_p=\beta \kappa b$~\cite{boal2012mechanics}. To relate the simulations to DNA~\cite{shin2012effects}, we identify one simulation unit of length with $10\,\mathrm{nm}$, so that $b=5\,\mathrm{nm}$. Since dsDNA has a persistence length of approximately $50\,\mathrm{nm}$, the simulated persistence length $\ell_p=5$ corresponds to the experimental value. This mapping sets the natural force scale to~\cite{shin2012effects}
\begin{equation}
    \frac{k_\mathrm{B}T}{\ell_p} \approx \frac{4.1\,\mathrm{pN\,nm}}{50\,\mathrm{nm}} \approx 80\,\mathrm{fN},
\end{equation}
so that $\beta f\ell_p=1$ corresponds to approximately $80\,\mathrm{fN}$. In particular, the force $f=0.3$ in simulation units, used in Fig.~\ref{fig:theory}b and Fig.~\ref{fig:SUP1_PRL}a, corresponds to $\beta f\ell_p=1.5$, or approximately $120\,\mathrm{fN}$, while the largest forces considered remain in the sub-piconewton regime relevant for the experiments of Chen \textit{et al.}~\cite{chen2010protein}. Equal and opposite forces were applied to the two terminal beads along the $z$-direction. For the WLC looping simulations, the interior contour length was chosen as $L=10.5$ in simulation units, corresponding to $105\,\mathrm{nm}$. This is close to the 305 bp looping segment ($\approx 305\times0.34\,\mathrm{nm} = 103.7\,\mathrm{nm}$) studied experimentally by Chen \textit{et al.}~\cite{chen2010protein}. Loop closure was defined by an absorbing boundary at $r_c=1$, corresponding to $10\,\mathrm{nm}$ in physical units. This value provides a reasonable estimate of the size of a protein complex capable of binding the two DNA binding sites.

\subsection{Simulation and sampling protocol} 

Simulations were initialized from straight chain conformations and equilibrated for $10^8$ timesteps using Langevin dynamics (\texttt{fix langevin}) with timestep $\Delta t=0.005$ and friction coefficient $\gamma=1$. Production runs subsequently employed the fast-forward Langevin integrator (\texttt{fix ffl})~\cite{hijazi2018fast} using the same timestep and friction coefficient. The determination of the force-dependent free energy and the handle projection $\langle\mathbf R_\mathrm{handles}\cdot\hat{\mathbf r}\rangle_{f=0;\mathbf r}$ (see next subsection) followed the procedure introduced in Refs.~\cite{laeremans2025theoretical, wout_laeremans_WLC}. For each force, $10^2$ independent production simulations of $10^9$ timesteps were performed. Every $10^3$ timesteps, the interior end-to-end distance was sampled. The resulting histogram was normalized to obtain the probability density $P(r)$, from which the free energy was calculated according to 
\begin{equation} 
    \beta F(r)=-\ln P(r). 
\end{equation} 
The same zero-force trajectories were used to compute $\langle\mathbf R_\mathrm{handles}\cdot\hat{\mathbf r}\rangle_{f=0;\mathbf r}$, from which the end-tangent alignment follows (see below). Looping times were determined following the methodology developed in Refs.~\cite{laeremans2024polymer, wout_laeremans_FJC}. Briefly, loop formation was monitored using the LAMMPS package REACTER~\cite{gissinger2020reacter}, in which a reaction was defined between the two terminal beads of the looping segment. This provides an efficient method to detect when the end-to-end distance falls below the capture radius $r_c$. The looping time was then defined as the corresponding first-passage time. For each force, $10^2$ independent simulations were performed and the reported looping times correspond to averages over these realizations.

\subsection{End-tangent alignment from zero-force simulations}\label{sec:SM_alignment}
The zero-force input of Eq.~\eqref{eq:handle} is the alignment $\langle\mathbf t_L\cdot\hat{\mathbf r}\rangle_{f=0;\mathbf r}$ between the end tangent of the interior segment and its end-to-end vector, shown in Fig.~\ref{fig:theory}a. We obtained it from the zero-force MD simulations with handles of length $h=\ell_p$. Consider a handle of $n=h/b$ bonds attached to a junction with tangent $\mathbf t$. At zero force, each bending angle $\theta$ between consecutive bonds is drawn independently from the weight $\sin\theta\,e^{-\beta\kappa\theta^2/2}$, with a uniformly distributed azimuth. On average, each bond therefore retains a fraction
\begin{equation}
c=\langle\cos\theta\rangle=\frac{\int_0^\pi d\theta\,\cos\theta\,\sin\theta\,e^{-\beta\kappa\theta^2/2}}{\int_0^\pi d\theta\,\sin\theta\,e^{-\beta\kappa\theta^2/2}}
\end{equation}
of the direction of the preceding bond, so that bond $k$ of the handle has mean orientation $c^k\mathbf t$. Summing over the handle bonds gives
\begin{equation}
\langle\mathbf R_h\rangle_{\mathbf t}=a_0\,\mathbf t,\qquad a_0=b\sum_{k=1}^{n}c^k=b\,c\,\frac{1-c^{n}}{1-c}. \label{eq:a0_SM}
\end{equation}
This is the discrete analogue of $a_0=\ell_p(1-e^{-h/\ell_p})$ for the continuous WLC. For $\kappa=10$, $b=0.5$ and $h=\ell_p$ ($n=10$), we find $c=0.906$ and $a_0=3.03$. Because the bending energy couples only neighbouring bonds, the handles depend on the interior segment only through its end tangents. Hence $\langle\mathbf R_1\rangle=a_0\mathbf t_0$ and $\langle\mathbf R_2\rangle=a_0\mathbf t_L$ for every interior conformation. Averaging over all conformations with interior end-to-end vector $\mathbf r$, projecting on $\hat{\mathbf r}$, and using $\langle\mathbf t_0\cdot\hat{\mathbf r}\rangle_{f=0;\mathbf r}=\langle\mathbf t_L\cdot\hat{\mathbf r}\rangle_{f=0;\mathbf r}$ (end-reversal symmetry) yields
\begin{equation}
\langle\mathbf t_L\cdot\hat{\mathbf r}\rangle_{f=0;\mathbf r}=\frac{\langle\mathbf R_\mathrm{handles}\cdot\hat{\mathbf r}\rangle_{f=0;\mathbf r}}{2a_0}. \label{eq:tL_from_R_SM}
\end{equation}
This relation is exact at zero force and holds for any handle length, provided $a_0$ is evaluated for that length. From the zero-force trajectories we sampled, every $10^3$ timesteps, the interior end-to-end vector $\mathbf r$ and the handle vector $\mathbf R_\mathrm{handles}$. The projection $\mathbf R_\mathrm{handles}\cdot\hat{\mathbf r}$ is binned in $r$ with the same bins as $P_0(r)$ and averaged within each bin. Dividing by $2a_0$ according to Eq.~\eqref{eq:tL_from_R_SM} gives the data in Fig.~\ref{fig:theory}a. As a consistency check, the alignment lies between $-1$ and $1$ and approaches unity as $r\to L$, where the interior segment is straight. The resulting $\langle\mathbf t_L\cdot\hat{\mathbf r}\rangle_{f=0;\mathbf r}$ is interpolated in $r$ and inserted in Eq.~\eqref{eq:handle}.

\subsection{Normalization of looping times} Figure~\ref{fig:experiment} compares experimental measurements, MD simulations, and theoretical predictions on a common scale. To remove the arbitrary overall timescale and focus on the force dependence, all looping times were normalized by their value at $f=60\,\mathrm{fN}$, the lowest experimental force reported by Chen \textit{et al.}~\cite{chen2010femtonewton,chen2010protein}, 
\begin{equation} 
    \tilde{\tau}(f) = \frac{\tau(f)} {\tau(60\,\mathrm{fN})}. 
\end{equation} 
For the continuous theoretical predictions, $\tau(60\,\mathrm{fN})$ was obtained directly from the calculated looping time curves. The experimental looping times were obtained as $\tau=1/k_L$ from the reported looping rates $k_L$. For Ref.~\cite{chen2010femtonewton} (60, 78, 121 and $183\,\mathrm{fN}$), the rates and their uncertainties were taken from Table~II of that work, and the uncertainties were propagated to $\tau$. For Ref.~\cite{chen2010protein}, the zero-noise value at $153\,\mathrm{fN}$ was taken from its supplementary rate table, and the values at $123$ and $175\,\mathrm{fN}$, which are not tabulated, were read off Fig.~4 of that work; no uncertainties are reported for these points. Since Ref.~\cite{chen2010protein} contains no zero-noise measurement at $60\,\mathrm{fN}$, both data sets were normalized by the $60\,\mathrm{fN}$ measurement of Ref.~\cite{chen2010femtonewton}. Force uncertainties are those reported in the original works. The simulations were not normalized at their own data point closest to $60\,\mathrm{fN}$, which would make the normalization sensitive to the statistical error of a single point. Instead, for both the handled ($h=\ell_p$ and $h=2\ell_p$) and no-handle ($h=0$) simulations, the looping times were first normalized by their own zero-force value and multiplied by the zero-force value of the corresponding theory (handle or no-handle), thereby removing the arbitrary overall timescale of the simulations. Theory and simulations were then divided by the same reference value, $\tau(60\,\mathrm{fN})$ of the corresponding theory curve. The handled simulations with $h=\ell_p$ and $h=2\ell_p$ are both compared with the same handle theory, since the handle effect has saturated by $h=\ell_p$. Uncertainties were propagated through the normalization by dividing the standard error of each looping time measurement by the corresponding normalization constant. Thus, if $\tilde{\tau}=\tau/\tau_\mathrm{ref}$, the associated uncertainty is 
\begin{equation} 
    \sigma_{\tilde{\tau}} = \frac{\sigma_\tau} {\tau_\mathrm{ref}}, 
\end{equation} 
where $\sigma_\tau$ denotes the standard error of the mean obtained from the independent simulation realizations.


\end{document}